\documentstyle[referee,psfig]{aa}
\begin{document}
\thesaurus{xxxxxxxx}
\title{X-ray Flux and Pulse Frequency Changes of Three High Mass X-ray 
Binary Pulsars: Vela X-1, GX 301-2 and OAO 1657-415} 
\titlerunning{Three High Mass X-ray Binary Pulsars}
\author{S{\i}tk{\i} \c{C}a\u{g}da\c{s} \.{I}nam \and Altan Baykal}
\authorrunning{\.{I}nam \& Baykal}
\institute{Department of Physics, Middle East Technical University, 06531 Ankara
, Turkey} 
\offprints{S.\c{C}.\.{I}nam}
\date{received / accepted}
\maketitle
\begin{abstract}
Using archival BATSE (Burst and Transient Source Experiment) 20-60 
keV band X-ray flux and pulse frequency time series, we look 
for correlations between torque, luminosity and specific angular momentum for 
three high mass X-ray binary pulsars Vela X-1, GX 301-2 and OAO 
1657-415. Our results show that there is no correlation between pulse 
frequency derivative and flux 
which may be an indication of the absence of stable prograde accretion 
disk. From the strong correlation of specific angular momentum and 
torque, we conclude that the accretion geometry changes continuously as 
suggested by the hydrodynamic simulations(Blondin et 
al. 1990).

\end{abstract} 
\section{Introduction}

Observations of accretion powered pulsars began with the discovery of
periodic X-ray pulsations from Cen X-3 by {\it{Uhuru}} (Giacconi et al.
1971; Schreier et al. 1972). Qualitative understanding of accretion
powered pulsars was achieved in the 1970s (Pringle\& Rees 1972; Davidson\&
Ostriker 1973; Lamb et al. 1973). Ghosh and Lamb presented an accretion 
disk theory to address the accretion powered pulsar observations in the 
1970s in terms of a fastness parameter, material and 
magnetic torques in the case of a stable prograde accretion 
disk (Ghosh\&Lamb, 1979a,b). In the absence of a stable accretion disk, 
numerical simulations were used to probe the nature of 
accretion (Anzer et al. 1987; Taam\&Fryxell 1988a,1988b,1989; Blondin et 
al. 1990).

Observations of pulse frequency changes in accretion powered pulsars are
direct signs of torques exerted on the pulsar. These torques can 
originate either outside or inside the star (Lamb et al. 
1978; Baykal \& \"Ogelman 1993). 
Internal torques depend on the coupling between interior components, in 
particular the core superfluid, and 
the solid outer crust (Baykal et al. 1991). External torques depend on the
magnetic field strength of the neutron star and on the type of accretion flow
to the neutron star. 

If the neutron star accretes mass from an accretion disk, torques are 
produced either  by the angular
momentum transfer of the plasma to the magnetic field in the
magnetospheric radius via interaction of the inner boundary of the disk
and the magnetic field lines (causing material torques) or by 
the interaction of the 
disk and the magnetic field (causing magnetic torques) (Ghosh \& Lamb 
1979a,b). If the accretion 
results from Roche lobe overflow of the companion, a persistent prograde 
Keplerian accretion disk forms and the disk creates material and magnetic 
torques causing the neutron star to spin-up or spin-down. For such a 
configuration, material torques can only give spin-up
contribution to the net torque, while magnetic torques may give either
spin-up or spin-down contribution.

If the companion does not fill its Roche lobe, then the neutron star
may still accrete mass from its companion's wind. From the hydrodynamic
simulations, it is seen that the stellar wind is
disrupted in the vicinity of a compact X-ray source (the neutron star for   
our case) which causes plasma to lose its homogeneity. The interaction of
the incident flow with the shock fronts around the neutron
star can produce retrograde and prograde temporary accretion disks 
(Anzer et al. 1987, Taam\& Fryxell 1988a, 1988b, 1989; Blondin et al. 
1990).  

The relations between X-ray luminosity, torque and specific angular 
momentum may lead to important clues about the accretion 
process. If the neutron star accretes mass from a stable prograde accretion 
disk, we expect a positive correlation between X-ray flux and 
torque (Ghosh\&Lamb 1979a,b). For the case of continuous changes in 
accretion 
geometry, we can expect a correlation between specific angular momentum 
and torque which may be the sign of significant torque changes while 
the luminosity does not vary significantly (Taam\&
Fryxell 1988a, 1988b, 1989; Blondin et al. 1990).      

In this paper, we use BATSE (Burst and Transient Source Experiment) 20-60 
keV 
band X-ray flux and pulse frequency time series of three high mass systems 
(Vela 
X-1, GX 301-2, and OAO 1657-415). This database is a part of the flux and 
pulse frequency database for accretion powered pulsars which was discussed 
before by Bildsten et al. (1997). 
Using these time series, we investigate the correlations of torque, X-ray 
luminosity and specific angular momentum. Detailed 
studies on torque and X-ray luminosity using the BATSE X-ray flux and pulse 
frequency data were 
presented before for GX 1+4 (Chakrabarty 1996,1997) and OAO 1657-415 
(Baykal 1997). Baykal(1997) also discussed correlations of specific 
angular momentum with torque and X-ray luminosity for OAO 1657-415.  
 
GX 1+4, which was continuously 
spinning-up in the 1970's, later exhibited a continuous spin-down 
trend with an anticorrelation of torque and X-ray luminosity, 
i.e the spin-down rate is increased with increasing X-ray luminosity 
(Chakrabarty 1996,1997). 
This spin-down episode was interpreted as evidence for a retrograde 
Keplerian 
accretion disk (Nelson et al. 1997) which may originate from the 
slow wind of a red giant (Murray et al. 1998). Other explanations for 
this spin-down episode were the radially advective 
sub-Keplerian disk (Yi et al. 1997) and warped disk (Van Kerkwijk et al. 
1998) models.

X-ray luminosity, torque and specific angular momentum correlations for 
OAO 1657-415 were studied earlier (Baykal, 1997). That work employed  
a flux and pulse frequency data string covering a $\sim$30\% shorter 
time interval compared to the content of the OAO 1657-415 data studied 
in the present paper. In that paper, correlations of pulse frequency 
derivative 
(proportional to torque exerted on the neutron star), pulse frequency 
derivative over flux 
(proportional to specific angular momentum of the accreted plasma) and flux 
(proportional to luminosity) were discussed. It was found that the most 
natural 
explanation of the observed X-ray flux and pulse frequency derivative 
fluctuations is 
the formation of temporary accretion disks in the case of stellar wind 
accretion. The present paper extends the analysis on OAO 1657-415 to 
cover a larger data string. We also present the results of a similar 
analysis in two other pulsars, Vela X-1 and GX 301-2. 

In the next section, database is introduced, 
and pulse frequency, pulse frequency derivative and flux time series are 
presented. A discussion of the 
results and conclusions are given in Section 3. 

\section{Database and Results}
BATSE is made up of eight detector modules located at the 
corners of CGRO (Compton Gamma Ray Observatory). These detectors have 
enabled continuous all sky monitoring
for both pulsed and unpulsed sources above 20 keV since 1991. BATSE
daily monitors the pulse frequency and X-ray flux of three low mass
binaries, five high mass binaries and seven previously known transients. It
has also discovered new transients (Bildsten et al. 1997).

This paper is based on orbitally corrected BATSE 20-60 keV band X-ray flux 
and pulse frequency 
time series of Vela X-1, GX 301-2 and OAO 1657-415 which are obtained from 
the ftp site
"ftp.cossc.gsfc.nasa.gov". In this paper, we assume that time variation 
of the 20-60 keV band flux represents the time variation of the 
bolometric X-ray flux.  It should be noted that the flux time series, 
reported in
Fig. 1, 4, and 7 might not be representative of the time variation of
the bolometric X-ray flux. Pulse 
frequency and flux time series 
are binned by considering that measurement errors dominate in short time 
lags or high frequencies in the pulse frequency derivative power spectrum 
(Bildsten et
al. 1997). We choose our bin sizes to the extent that the measurement
errors do not dominate on pulse frequency derivatives. Bin sizes are 45 days, 
30 days and 16 days for Vela X-1, GX 301-2 and OAO 1657-415 respectively. 
Pulse frequency derivatives are found by 
averaging left and 
right derivatives of pulse frequency values so that each pulse frequency 
derivative 
corresponds to a single flux and time value. We present X-ray flux, pulse 
frequency and pulse frequency derivative time series in Fig. 1,4 and 7. In 
Fig. 
2, 5 and 8, we present the plot of pulse frequency derivative and flux 
values 
corresponding to the same time value. Since pulse frequency derivative and 
X-ray flux are directly proportional to torque and X-ray luminosity, these 
figures show the relation between torque and X-ray luminosity. Pulse 
frequency derivative over flux is 
directly proportional to the specific angular momentum of the plasma 
($l=I\dot{\Omega}/\dot{M}$ where l is the specific angular momentum, 
$\dot{\Omega}$ is the frequency derivative and $\dot{M}$ is the mass 
accretion rate).
Fig. 3, 6, 9 which are plots of pulse frequency derivative over flux and 
pulse frequency derivative show the relation between specific angular 
momentum and torque. Pulse frequency derivative over flux time series are 
created by dividing each pulse frequency derivative value with the flux 
value corresponding to the same time.

\subsection{Vela X-1}
283s pulsations from Vela X-1 were discovered by \it SAS-3 \rm in 1975
(McClintock et al. 1976). It is the brightest
persistent accretion powered pulsar in the 20-60 keV energy band  
(Bildsten et al. 1997). Optical companion of Vela X-1 is the B0.5 Ib
supergiant HD77581 (Vidal et al. 1973). This system is an eclipsing
binary with eccentricity of $\simeq 0.126$ and period of 8.96days (Rappaport 
et al. 1976).

X-ray flux, pulse frequency and pulse frequency derivative values of Vela 
X-1 cover the interval 
between 48371 MJD and 50580 MJD (Fig.1). In Fig.2, pulse frequency 
derivative and corresponding X-ray flux data were plotted. No correlation 
between frequency
derivative and flux is found. From Fig.3, it is seen that there is a
correlation between pulse frequency derivative over flux and frequency
derivative.

\begin{figure}
\begin{center}
\psfig{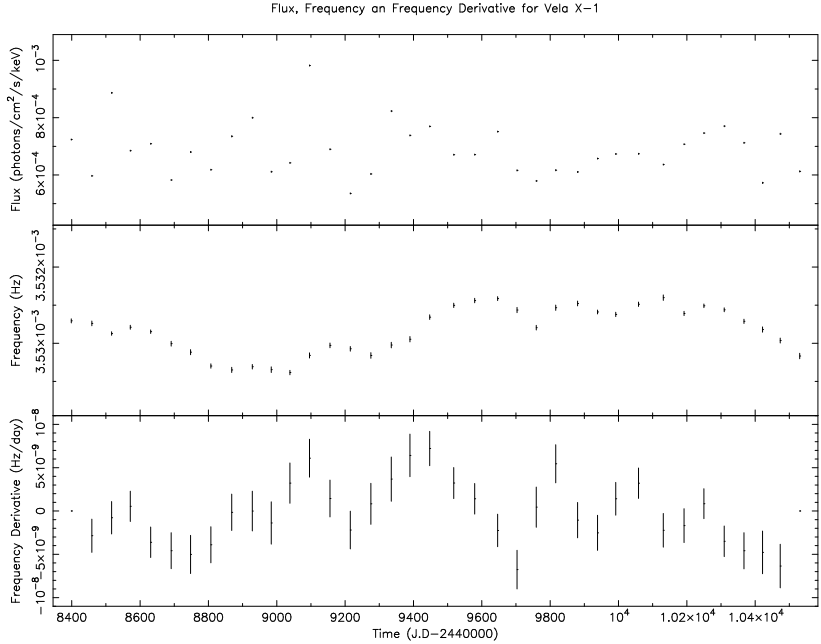}  
\end{center}
\caption{Flux, Pulse Frequency and Pulse Frequency Derivative Time Series of 
Vela X-1. Data points were obtained by making bins each covering 45 days 
from the original data. Errors are in 1$\sigma$ level.} 
\end{figure}

\begin{figure}
\begin{center}
\psfig{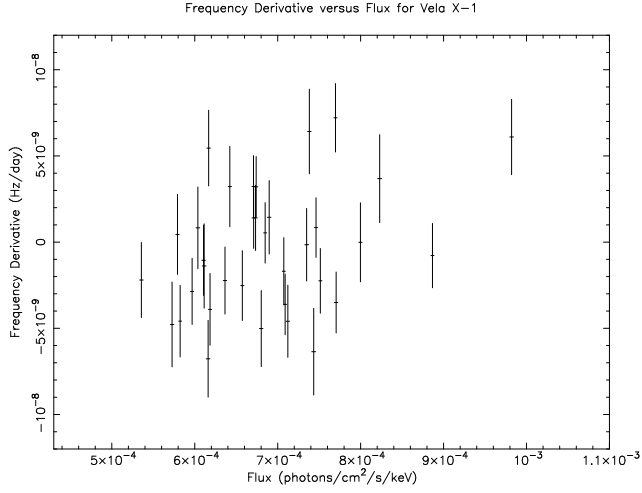}
\end{center}
\caption{Pulse Frequency Derivative versus Flux for Vela X-1. Errors are 
in 1$\sigma$ level.} \end{figure}  

\begin{figure}
\begin{center}
\psfig{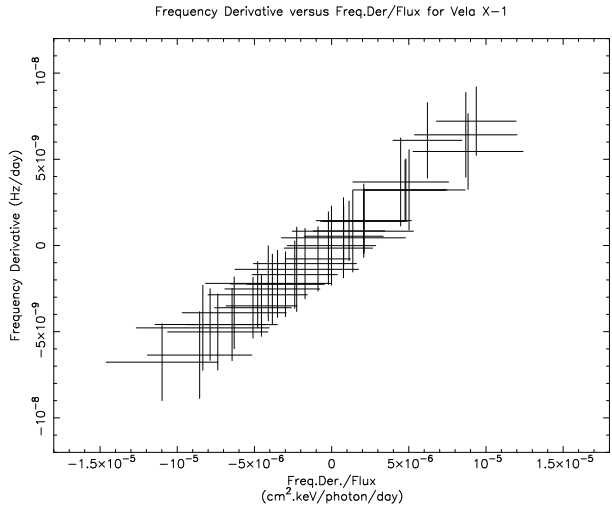}
\end{center}
\caption{Pulse Frequency Derivative versus Pulse Frequency derivative over 
Flux for Vela X-1. Errors are in 1$\sigma$ level.}
\end{figure}

\subsection{GX 301-2}
700s pulsations from GX 301-2 (4U 1223-62) were discovered by Ariel 5 in
1975 (White et al. 1976). GX 301-2  was, on
average, neither spinning up nor spinning down between 1975 and 1985. 
After 1985, a spin-up episode began
(Nagase 1989), reaching the current pulsar spin period of $\sim 
676s$. GX
301-2, being in a 41.5 day eccentric orbit ($e=0.47$), is an accreting
pulsar with the supergiant companion Wray 977 (Sato et al. 1986). This
source exhibited two rapid spin-up episodes at $\sim 48450$
MJD and $\sim
49250$ MJD. These two spin-up
episodes suggest the existence of a long-lived ($\approx 30$days) accretion
disk (Koh et al. 1997).

X-ray flux, pulse frequency and pulse frequency derivative 
values of GX 
301-2 cover the interval between 48371 MJD and 50577 MJD (Fig.4). Fig.5 
is the plot of pulse frequency derivative and corresponding X-ray flux
values. No correlation between
pulse frequency derivative and flux is found. From Fig.6, we
see that there is a
correlation between pulse frequency derivative over flux and pulse 
frequency
derivative. In Fig. 5 and 6, there exist two points with pulse frequency
derivative values greater than $10^{-7}$ Hz/day. These points correspond
to two rapid spin-up episodes which was interpreted as a sign of
transient prograde accretion disk (Koh et al. 1997).  

\begin{figure}
\begin{center}
\psfig{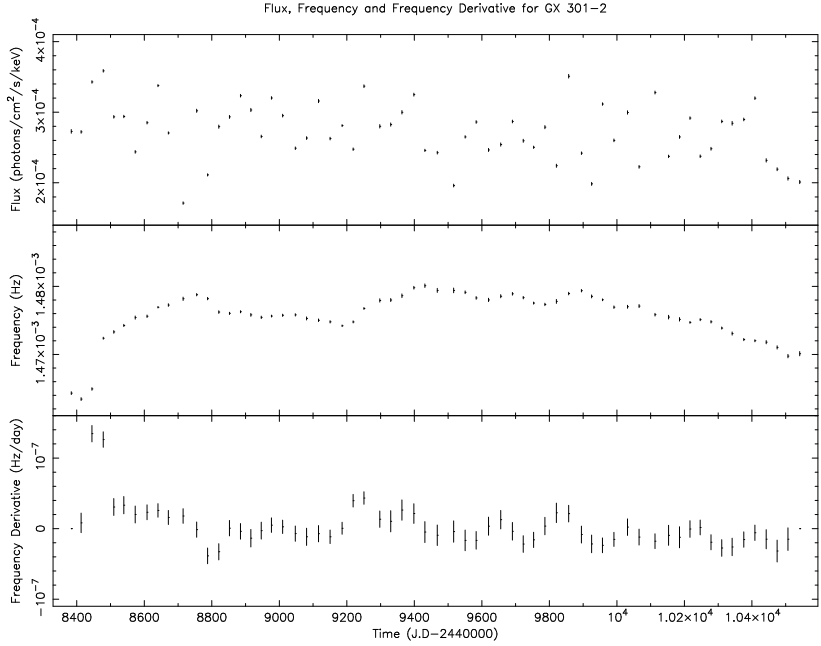}
\end{center}
\caption{Flux, Pulse Frequency and Pulse Frequency Derivative Time Series 
of GX 
301-2. Data points were obtained by making bins each covering 30 days
from the original data. Errors are in 1$\sigma$ level.}
\end{figure}

\begin{figure}
\begin{center}
\psfig{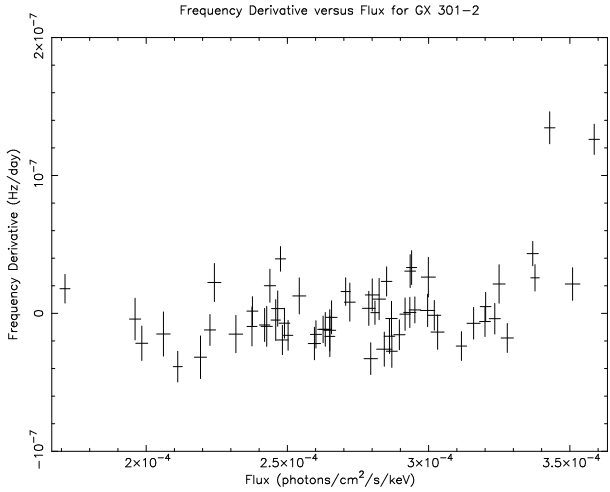}
\end{center}
\caption{Pulse Frequency derivative versus Flux for GX 301-2. Two points 
with 
pulse frequency derivative values greater than $10^{-7}$ Hz/day, 
correspond to two rapid spin-up episodes discussed by Koh et al. (1997). 
Errors are in 1$\sigma$ level.} \end{figure}

\begin{figure}
\begin{center}
\psfig{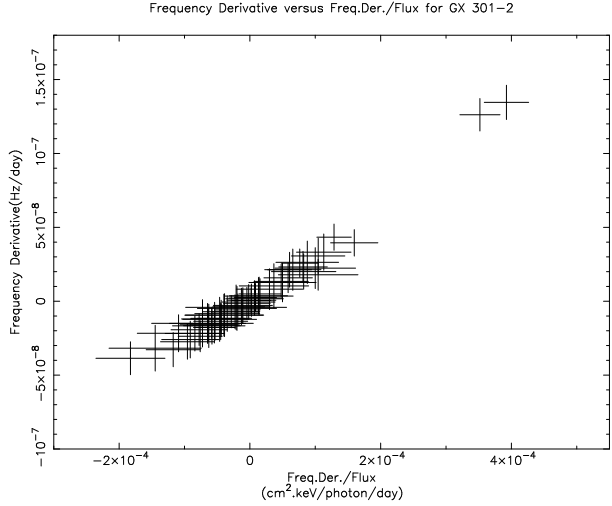}
\end{center}
\caption{Pulse Frequency Derivative versus Pulse Frequency Derivative over 
Flux for GX 301-2.Two points with 
pulse frequency derivative values greater than $10^{-7}$ Hz/day, correspond 
to two rapid spin-up episodes discussed by Koh et al. (1997). Errors are 
in 1$\sigma$ level.} \end{figure}

\subsection{OAO 1657-415}
OAO 1657-415 was first detected by the Copernicus satellite (Poldan et 
al. 1978). 38.22s pulsations from OAO 1657-415 were found in 1978 from 
HEAO 1 observations (White et al. 1979). The optical companion of OAO 
1657-415 is probably a OB type star. Its binary orbit period with an 
eccentricity of $\sim 0.10$  
was found to be 10.4 days from the eclipse due to its companion from 
timing observations of this source with the BATSE observations (Chakrabarty 
1993). Detailed studies on X-ray flux and pulse frequency derivative changes 
were performed earlier on a less extensive  
BATSE 20-60keV X-ray flux and pulse frequency data string (Baykal 1997). 
In the current data string, X-ray flux and pulse frequency and 
pulse frequency derivative values of OAO 1657-415 
cover the interval between 48372 MJD and 50302 MJD (Fig.7).

\begin{figure}
\begin{center}
\psfig{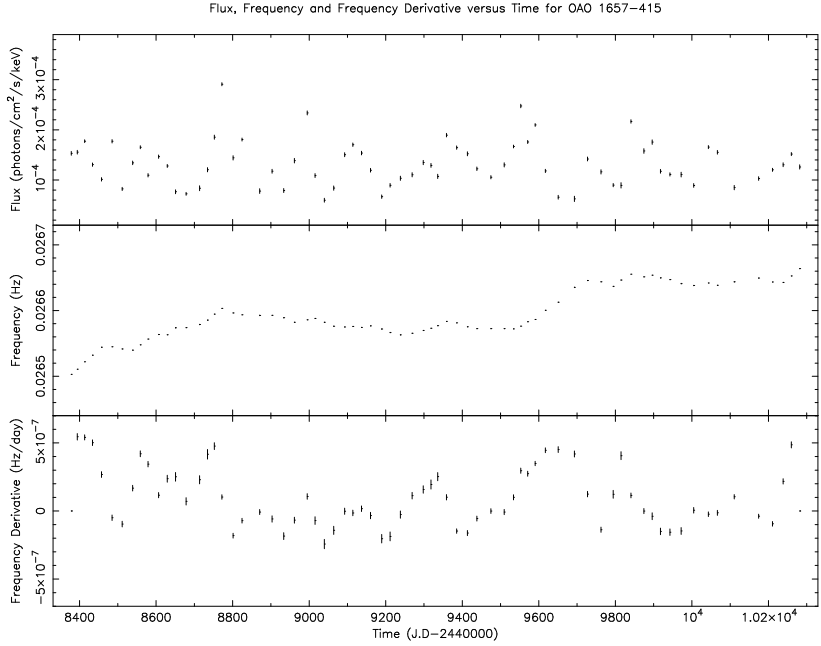}
\end{center}  
\caption{Flux, Pulse Frequency and Pulse Frequency Derivative Time Series of 
OAO 
1657-415. Data points were obtained by making bins each covering 16 days
from the original data. Errors are in 1$\sigma$ level.}
\end{figure}

Fig. 8 presents pulse frequency derivative and corresponding X-ray flux 
time series. No correlation between pulse frequency
derivative and flux is found. There is a
correlation between pulse frequency derivative over flux and pulse frequency 
derivative as seen from Fig. 9.

\begin{figure}
\begin{center}
\psfig{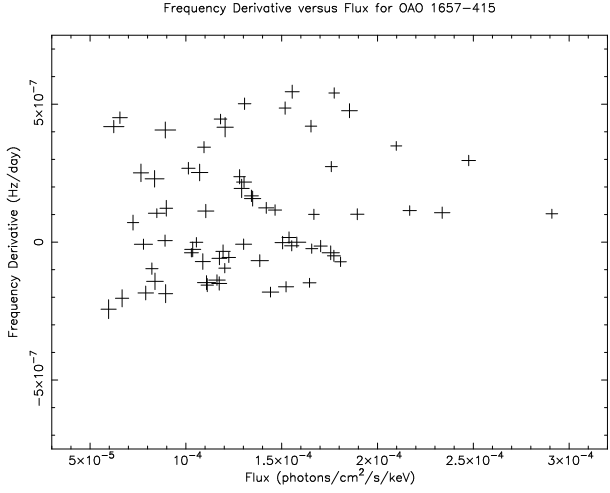}
\end{center}
\caption{Pulse Frequency Derivative versus Flux for OAO 1657-415. Data points
were obtained by making bins each covering 16 days
from the original data. Errors are in 1$\sigma$ level.}
\end{figure}

\begin{figure}
\begin{center}
\psfig{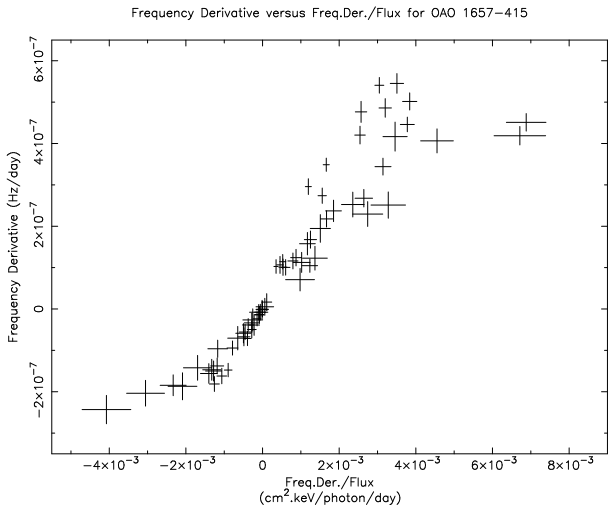}
\end{center}
\caption{Pulse Frequency Derivative versus Pulse Frequency Derivative over 
Flux for OAO 1657-415. Errors are in 1$\sigma$ level.}
\end{figure}

\section{Discussion and Conclusion}
The torque on the neutron star can be expressed in terms of the specific
angular momentum ($l$) added to the neutron star by the accreted plasma
and the mass accretion rate ($\dot{M}$):

\begin{equation}
N=I\dot{\Omega}=\dot{M}l.
\end{equation}

where $\dot{\Omega}$ is the spin frequency derivative and $I$ is the 
moment of inertia of the neutron star.   

This equation is a general expression which is valid for both accretion
from a Keplerian disk and accretion from the stellar wind. In the case of 
accretion from a Keplerian disk (Ghosh\& Lamb 1979b) we have,

\begin{equation}
I\dot{\Omega}=n(\omega_s)\dot{M}l,
\end{equation}

and

\begin{equation}
l=(GMr_{0})^{1/2},
\end{equation}

 where $\omega_s$ is the fastness parameter and $n(\omega_s)$,
the dimensionless torque, represents the
ratio of the 
total (magnetic plus material) torque to the material torque. $r_0$ 
is the radius of the inner edge of the disk, which can be written as 
$r_0\simeq 0.5(2GM)^{-1/7}\mu^{4/7}\dot{M}^{-2/7}$ where $\mu$ is the 
magnetic moment of the neutron star. Dimensionless torque can 
approximately be written as 

\begin{equation}
n(\omega_s)=1.4{{1-\omega_s/\omega_c} \over {1-\omega_s}},
\end{equation}

where $\omega_c$ is a critical value for the fastness parameter 
which defines the boundary between spin-up and net 
spin-down phases. Dimensionless torque may be positive or negative 
depending on the fastness parameter $\omega _{s}$ which is defined as

\begin{equation}
\omega_s={{\Omega_s} \over {\Omega_{K_0}}},
\end{equation}

where $\Omega_s$ is the neutron star's spin angular velocity and 
$\Omega_{K_0}$ is the angular velocity corresponding to the Keplerian 
velocity at the magnetospheric radius. For a slowly rotating
neutron star for which $\omega _s < \omega _c\simeq 0.35-0.95$ 
(Ghosh\&Lamb 1979b; Wang 1995; Li\&Wang 1996,1999), we expect a 
spin-up torque and for
a very fast rotating neutron star ($\omega _s >> \omega _c$), we expect a
spin-down torque. We also expect to see positive correlation between
torque and mass accretion rate if the disk is prograde. For a disk formed
from Roche Lobe overflow of the companion we expect the plasma to carry
positive specific angular momentum, so a prograde disk should be 
formed. The total torque exerted on the neutron star is  
proportional to the material torque for a given fastness parameter. Since 
the specific angular momentum weakly depends on mass accretion rate 
($l\propto \dot{M}^{-1/7}$), the net 
torque becomes proportional to the mass accretion rate ($\dot{M}$). The 
bolometric X-ray luminosity is also correlated with the mass 
accretion rate ($L=GM\dot{M}/R$ where R is the radius of the neutron star). 
Thus, we expect a correlation between 
torque and X-ray luminosity for the sources accreting from prograde 
accretion disks (torques are positive). For the similar reasons, an 
anticorrelation between torque and X-ray luminosity is expected from a 
retrograde accretion disk (torques are negative). However, for the 
pulsars we have considered, we see no correlation of pulse frequency 
derivative 
and flux(Fig. 2,5,8). Moreover, there are several transitions from 
spin-up and spin-down. These results 
suggest that we do not have stable accretion disks for these sources. 

There is a model which does not exclude the possibility of a stable 
Keplerian accretion disk(Anzer\&B\"orner 1995). This model is proposed to 
explain the torque reversals in Vela X-1 suggesting the existence of a 
stable accretion disk lying just outside the magnetosphere mass of which 
exhibits small variations. It is found that change of the disk's mass in 
a random way can produce variations of torque in Vela X-1. However, the 
authors have not identified a specific physical process which is 
responsible for such random mass fluctuations. 

Alternatively, we can think of the existence of accretion geometry 
changes around 
the neutron star. We can, in general, write the variation of torque 
($\delta{N}$) as

\begin{equation}
\delta{N}=I\delta\dot{\Omega}=\delta\dot{M}l+\dot{M}\delta l
\end{equation}

where I is the moment of inertia of the neutron star, l is the specific 
angular momentum of the accreting matter, $\delta\dot{\Omega}$, 
$\delta\dot{M}$, and 
$\delta l$ are the variations of spin frequency derivative, mass accretion 
rate (proportional to X-ray flux) and
specific angular momentum (proportional to pulse frequency derivative over 
flux) respectively. When the variations of torques 
for the three pulsars are concerned, we have considerable changes 
and transitions from negative values to positive values and vice versa. 
For the case of an accretion from the wind, numerical simulations show
that the changes in the sign of specific angular momentum is possible
(Anzer et al. 1987; Taam\& Fryxell 1988a,1988b,1989; Blondin et al. 1991, 
Murray et al.
1998). So, we can observe transitions from spin-up to spin-down or vice 
versa even if there is not a significant change in mass accretion rate. 
For wind accreting sources, continuos change in accretion geometry rules out 
the existence of a 
stable accretion disk. Thus, it is unlikely to see a correlation between 
torque and X-ray luminosity which is the case for our sources as well. For 
such sources, a correlation between specific angular 
momentum and torque shows that there are considerable changes in torque 
while there are not very considerable changes in X-ray luminosity. This 
indicates changes in accretion geometry. The correlation between 
specific angular momentum and torque exists for all of the three sources 
(Fig. 3, 6 and 9). 

There are recent developments which explain the torque reversals in 
accretion powered pulsars. Negative 
torques may come from a retrograde Keplerian accretion disk (Nelson et 
al. 1997) which may, for instance, originate from a red giant 
(Murray et al. 1998). These spin-down torques may be the result of an 
advection dominated sub-Keplerian disk for which the fastness parameter 
should be higher than that of a corresponding Keplerian disk causing a 
net spin-down (Yi et al. 1997) or the warping of the disk so that the 
inner disk is tilted by more than 90 degrees (van Kerkwijk et al. 1998). 
Torque and X-ray luminosity correlation is expected from these 
models which is not found for our sources. 
Moreover, in these models timescales for torque reversals are either not 
certain or of the 
order of years. So, these ideas about torque reversals are not supported 
by the behaviour of the three pulsars we have considered.  

Our considerations about correlations between pulse frequency derivative, 
flux and 
specific angular momentum give insights about the physics of the plasma 
flow in the vicinity of the neutron stars accreting from the winds' of 
their companions. We had similar conclusions for all three sources. 
We found that it is 
unlikely for these sources to have stable prograde Keplerian accretion 
disks since they show both spin-up and spin-down episodes and they do not 
show correlation between torque and luminosity. Correlation between specific 
angular momentum and torque for these 
sources may indicate the continuous change in accretion geometry. This 
shows the possibility of temporary prograde and retrograde accretion disk 
formation. It would be also possible to have stronger idea about the 
accretion geometry if we had measurements of the pulse frequency derivative 
with a time resolution of the order of hours, 
which is a typical time scale of accretion geometry changes for these 
systems (Taam\& Fryxell 1988a, 1988b, 1989; Blondin et al. 
1990). For such a 
case, we would allow a better comparison of the changes in flux and 
specific angular momentum and it would be interesting to detect both  
high flux and 
low specific angular momentum points corresponding to the radial flow cases 
and to see low flux and high specific angular momentum points 
corresponding to the accretion from prograde (high positive specific 
angular momentum) or retrograde (high negative specific angular momentum) 
accretion disks.

Our flux and pulse frequency time series are found  using the X-ray flux  
and pulse frequency between 
20-60 keV. Observatories with higher time resolution and capable of 
detecting photons from a wider energy band will be useful to observe the 
torque, 
X-ray luminosity and specific angular momentum changes for these pulsars. 
An encouraging example is given by RXTE (Rossi X-ray Timing Explorer) 
observations of the accretion powered 
pulsar 4U 1907+09. Dipping 
activity in the X-ray intensity was found which was interpreted as a 
consequence of inhomogeneity of the wind from the companion(In'T Zand et 
al. 1997), and more recently, it was shown from the RXTE observations 
that 4U 
1907+09 exhibited transient $\sim 18$s QPO oscillations during a flare 
which was superposed on long term spin-down rate. This was interpreted as a 
sign of 
transient retrograde accretion disk (In'T Zand et al. 1998a,b). 
RXTE observations of Vela X-1, GX 301-2, and OAO 1657-415 may 
provide more understanding on the nature of accretion flow through tests 
on flux and frequency time series at higher resolution.

\begin{acknowledgements}
We acknowledge Dr.Ali Alpar and Dr.\c{S}\"olen Balman for critical 
reading of the manuscript.
It is a pleasure to thank Dr.Matthew Scott and Dr.Bob Wilson for their 
help to our questions about 
the database. We thank the Compton Gamma Ray Observatory team at HEASARC 
for the archival data.
\end{acknowledgements}

\end{document}